# Electronic Structure and the Properties of Phosphorene Systems

Shuhei Fukuoka, Toshihiro Taen, and Toshihito Osada*

*Institute for Solid State Physics, University of Tokyo,*

*5-1-5 Kashiwanoha, Kashiwa, Chiba 277-8581, Japan.*

A single atomic layer of black phosphorus, phosphorene, was experimentally realized in 2014. It has a puckered honeycomb lattice structure and a semiconducting electronic structure. In the first part of this paper, we use a simple LCAO model, and discuss qualitatively the electronic structure of phosphorene systems under electric and magnetic fields, especially, noting their midgap edge states. The next part is spent for the review of the research progress on phosphorene in the past one year since its appearance in 2014. Phosphorene has been a typical object to study the semiconductor physics in atomic layers.

## 1. Introduction: from black phosphorus to phosphorene

Phosphorene, an atomic layer of phosphorus, was realized by the mechanical exfoliation from the layered crystal of black phosphorus (BP) in 2014.[1] It is a single component two-dimensional (2D) crystal following graphene and silicene. In contrast to graphene, which is a 2D zerogap conductor, phosphorene is a 2D semiconductor with a finite band gap. Because of its high hole mobility, phosphorene has been expected to be a hopeful *p*-type device material, which is complementary to the n-type atomic layers like MoS$_2$.[1-3]

The black phosphorus (BP) has the most stable crystal structure at ambient pressure among the several allotropes of phosphorus. It is formed by alternative stacking of phosphorene layers, in which phosphorus atoms form a puckered honeycomb lattice. This structure is different from the A7 structure of other pnictogen (group 15) crystals. Since BP was the third elemental semiconductor following germanium and silicon, band calculations using tight-binding and pseudo-potential methods[4-6] and experimental studies using high-quality single crystals[7-9] were intensively performed mainly in 1980s. BP is a p-type narrow gap semiconductor with the direct gap of 300 meV. Although the interlayer conductivity is much smaller than the in-plane conductivity reflecting the layered structure, the interlayer coupling is not negligible at the ambient pressure. The in-plane conductivity also has anisotropy reflecting the puckered honeycomb structure of each layer. The temperature dependence of resistance shows the behavior characteristic to the narrow gap semiconductors in which both electrons and holes contribute to transport at high temperature. The undoped BP is a p-type semiconductor, but the origin of natural p-type doping has been unclear. Under the pressure, BP undergoes the structural phase transition into the semimetal phase with the A7 structure at ~ 1.5 GPa, and to the metallic phase, which shows the superconductivity, with the simple cubic structure at ~ 10 GPa.

In 2014, Liu et al. extracted monolayer phosphorene from the layered BP crystal by employing the mechanical exfoliation technique, and demonstrated its semiconducting transport properties.[1] After this report, a huge amount of research has been performed on its physical properties and possible applications.

In this paper, we review the research progress on phosphorene in the past one year since its appearance. Before the review, first, we take a qualitative view on the electronic structure of phosphorene systems using a simple LCAO model.

## 2. LCAO model for phosphorene systems

In advance of the review, we study the qualitative electronic structure of monolayer and multilayer phosphorene under vertical electric and magnetic fields using a simplified LCAO model.[10] Our model is based on the Slater-Koster-Harrison method.[11,12] This tight-binding approach allows us to find characteristic features of the electronic structure easily, although it has no quantitative accuracy. The crystal structure of monolayer phosphorene is a puckered honeycomb lattice, where phosphorus atoms exist on two parallel planes as shown in Fig. 1(a) and (b). So, even in the monolayer, the external vertical electric field can modify the electronic structure due to the potential difference between the two planes. In the same way, the parallel magnetic field can affect electronic structure since the single layer has finite thickness which the magnetic flux can penetrate. Experimentally, the vertical electric field can be applied using the FET device with dual gate electrodes (top gate and bottom gate).

## 2.1 LCAO model for phosphorene based on Slater-Koster-Harrion method

First, we consider monolayer phosphorene under the electric field. The matrix element (transfer integral) between adjacent two atoms is approximately given by $V_{ll'm}(d) = \eta_{ll'm}\hbar^2/m_0 d^2$, where $d$ and $m_0$ are the inter-atomic distance and the electron rest mass, respectively. $l$ and $l'$ denote the orbital quantum numbers ($s$, $p$) of the two atoms, and $m$ is the common magnetic quantum number ($\sigma$, $\pi$) along the axis connecting two atoms. We adopt the adjusted values, $\eta_{ss\sigma} = -1.40$, $\eta_{sp\sigma} = 1.84$, $\eta_{pp\sigma} = 3.24$, and $\eta_{pp\pi} = -0.81$ given by Harrison.[12] The normal electric field introduces the potential difference $2\Delta$ between two planes. The unit cell contains four atoms labeled A, B, A', and B' in Fig. 1(a) and (b). Their positions are presented as $\boldsymbol{\tau}_A = (uc, 0, vb)$, $\boldsymbol{\tau}_B = ((1/2 - u)c, a/2, vb)$, $\boldsymbol{\tau}_{A'} = -\boldsymbol{\tau}_A$, and $\boldsymbol{\tau}_{B'} = -\boldsymbol{\tau}_B$. We use the lattice parameters of BP for phoshorene:[4] $c = 4.376\,\text{Å}$, $a = 3.314\,\text{Å}$, $b = 10.478\,\text{Å}$, $u = 0.08056$, and $v = 0.10168$. The displacement vectors are defined by $\boldsymbol{d}^{(1)} = \boldsymbol{\tau}_B - \boldsymbol{\tau}_A$, $\boldsymbol{d}^{(1)'} = (\boldsymbol{\tau}_B - \boldsymbol{a}) - \boldsymbol{\tau}_A$, $\boldsymbol{d}^{(2)} = \boldsymbol{\tau}_{A'} - \boldsymbol{\tau}_A$, and $\bar{\boldsymbol{d}}^{(2)} = \boldsymbol{\tau}_{B'} - \boldsymbol{\tau}_B$ with $\boldsymbol{a} = (0, a, 0)$ and $\boldsymbol{c} = (c, 0, 0)$.

The tight-binding Hamiltonian is represented as the $16 \times 16$ matrix, in which the Bloch sums of $3s$, $3p_x$, $3p_y$, and $3p_z$ orbitals at A, B, A', and B' sites are taken as the bases.

$$H_{\text{mono}}(\boldsymbol{k}; \Delta) = \begin{pmatrix} M_0 + \Delta & M_1 & M_2^+ & 0 \\ {}^tM_1^* & M_0 + \Delta & 0 & M_2^- \\ {}^tM_2^{+*} & 0 & M_0 - \Delta & {}^tM_1^* \\ 0 & {}^tM_2^{-*} & M_1 & M_0 - \Delta \end{pmatrix}. \quad (1)$$

The diagonal $4 \times 4$ matrices express the energies of four atomic sites with

$$M_0 = \begin{pmatrix} \varepsilon_s & 0 & 0 & 0 \\ 0 & \varepsilon_p & 0 & 0 \\ 0 & 0 & \varepsilon_p & 0 \\ 0 & 0 & 0 & \varepsilon_{]p} \end{pmatrix}. \tag{2}$$

Here, $\varepsilon_s = -17.10$ eV and $\varepsilon_p = -8.33$ eV are the energy levels of $3s$ and $3p$ orbitals of phosphorus, respectively.[12] The nearest neighbor coupling between A and B (B' and A') and the next nearest neighbor coupling between A and A' (B and B') are represented by the matrices $M_1$ and $M_2^{\pm}$, respectively.

$$M_1 = \begin{pmatrix} E_{ss}^{(1)} g_1^+ & E_{sx}^{(1)} g_1^+ & E_{sy}^{(1)} g_1^- & 0 \\ -E_{sx}^{(1)} g_1^+ & E_{xx}^{(1)} g_1^+ & E_{xy}^{(1)} g_1^- & 0 \\ -E_{sy}^{(1)} g_1^- & E_{xy}^{(1)} g_1^- & E_{yy}^{(1)} g_1^+ & 0 \\ 0 & 0 & 0 & E_{zz}^{(1)} g_1^+ \end{pmatrix}, \tag{3}$$

$$M_2^{\pm} = \begin{pmatrix} E_{ss}^{(2)} g_2^{\pm} & \pm E_{sx}^{(2)} g_2^{\pm} & 0 & E_{sz}^{(2)} g_2^{\pm} \\ \mp E_{sx}^{(2)} g_2^{\pm} & E_{xx}^{(2)} g_2^{\pm} & 0 & \pm E_{sz}^{(2)} g_2^{\pm} \\ 0 & 0 & E_{yy}^{(2)} g_2^{\pm} & 0 \\ -E_{sz}^{(2)} g_2^{\pm} & \pm E_{xz}^{(2)} g_2^{\pm} & 0 & E_{zz}^{(2)} g_2^{\pm} \end{pmatrix}. \tag{4}$$

Here, $E_{ss}^{(i)} = V_{ss\sigma}(d^{(i)})$, $E_{s\alpha}^{(i)} = (d_\alpha^{(i)}/d^{(i)}) V_{sp\sigma}(d^{(i)})$, and $E_{\alpha\beta}^{(i)} = \left(d_\alpha^{(i)} d_\beta^{(i)} / d^{(i)2}\right) V_{pp\sigma}(d^{(i)}) + \left(\delta_{\alpha\beta} - d_\alpha^{(i)} d_\beta^{(i)} / d^{(i)2}\right) V_{pp\pi}(d^{(i)})$ are the Slater-Koster inter-atomic matrix elements,[11] where $\boldsymbol{d}^{(i)} = \left(d_x^{(i)}, d_y^{(i)}, d_z^{(i)}\right)$ and $d^{(i)} = |\boldsymbol{d}^{(i)}|$ ($i = 1, 2$ and $\alpha, \beta = x, y, z$). The phase factors are defined as $g_1^{\pm} = \exp(i\boldsymbol{k} \cdot \boldsymbol{d}^{(1)}) \pm \exp(i\boldsymbol{k} \cdot \boldsymbol{d}^{(1)'})$, $g_2^+ = \exp(i\boldsymbol{k} \cdot \boldsymbol{d}^{(2)})$, and $g_2^- = \exp(i\boldsymbol{k} \cdot \overline{\boldsymbol{d}}^{(2)})$.

*2.2 Band structure of monolayer and multilayer phosphorene*

The Hamiltonian $H_{\text{mono}}(\boldsymbol{k}; \Delta)$ gives the band dispersion of monolayer phosphorene as shown in Fig. 2(a) and Fig. 3(a). We can see that sixteen bands are divided into eight pairs. The lowest two pairs, three pairs just below the gap, and three pairs above the gap can be roughly regarded as the $3s$ bands, the $3p$ bonding bands, and the $3p$ anti-bonding bands, respectively. Each pair degenerate at the zone boundary (XM and YM line) reflecting the equivalence of A and A' (B and B') atoms in the projected 2D lattice. A large direct gap opens at the Γ point ($\boldsymbol{k} = \boldsymbol{0}$) between the $3p$ bonding and anti-bonding band sets. Moreover, we can see that the dispersion of the valence band is anisotropic around the Γ point. The hole mass along the armchair (ΓX) direction is lighter than that along the zigzag (ΓY) direction, resulting in the transport anisotropy, that is, higher conductivity along the armchair ($x$) direction. At several points, the neighboring band pairs connect with each other forming the Dirac cones with the linear dispersion, reflecting a modified honeycomb lattice of 2D projection similar to graphene.

Once the electric field becomes finite (Δ≠ 0), the degeneracy along the XM boundary breaks, and the energy gap decreases reflecting the increase of band width as shown in Fig 3(b). The dispersion anisotropy is also affected by the electric field.

By taking the direct sum of $H_{\text{mono}}(\boldsymbol{k}; \Delta)$ and setting the interlayer coupling matrices, the Hamiltnian for multilayer phosphorene can be easily constructed. Here, we take into account only the nearest neighbor interlayer coupling between A' and C (B' and D) in Fig.1(c). The band structure of bilayer phosphorene is shown in Fig. 2(b) and Fig. 3(c). The main gap considerably decreases from the monolayer. For larger layer numbers, the main gap becomes much smaller, for example, as seen in Fig. 2(c). Above features mean that the band gap of phosphorene systems can be controlled by the layer number and the vertical electric field.

*2.3 Edge state in monolayer and bilayer phosphorene*

In the phosphorene with finite size, the edge state appears in the main gap. It can be investigated by considering the phosphorene nanoribbon with finite width in one direction. In the same way as multilayers, its Hamiltonian can be constructed from that of the single chain with the width of one unit cell. At the zigzag edge in monolayer, one edge subband appears in the middle of main gap. It is isolated from both of the conduction and valence bands, and doubly degenerated corresponding to two edges of the nanoribbon. This degeneracy is broken under the vertical magnetic field, if the zigzag edges of two sides of the nanoribbon lie on different atomic planes of monolayer. At the armchair edge, a pair of edge subbands with two-fold degeneracy appear because the armchair edge consists of the upper and lower plane sites. In the bilayer nanoribbon, a pair of edge subbands appear corresponding to the bonding and anti-bonding state of the edge states of two monolayers. These edge subbands overlap with the valence band in energy, so that the charge transfer occurs between the edge states and the valence band. In other words, the edge state could work as acceptor. This might cause the natural hole-doping around the edge of the bilayer phosphorene.

*2.4 Aharonov-Bohm effect in monolayer phosphorene under the in-plane magnetic field*

Although the 2D band structure of phosphorene is not quantized into the Landau levels under the in-plane magnetic field, it is modified since the puckered lattice of monolayer has finite thickness. We consider the monolayer phosphorene under the in-plane magnetic field perpendicular to the zigzag edge, $\mathbf{B} = (B, 0, 0)$. Choosing the gauge as $\mathbf{A} = (0, -Bz, 0)$, the Hamiltonian is written by

$$H_{\text{mono}}(\mathbf{k} + (e/\hbar)\mathbf{A}; \Delta) = \begin{pmatrix} M_0 + \Delta & M_{1u} & M_2^+ & 0 \\ {}^t M_{1u}{}^* & M_0 + \Delta & 0 & M_2^- \\ {}^t M_2^{+*} & 0 & M_0 - \Delta & {}^t M_{1d}{}^* \\ 0 & {}^t M_2^{-*} & M_{1d} & M_0 - \Delta \end{pmatrix}. \quad (5)$$

Here, $M_{1u}$ and $M_{1d}$ are given by substituting the phase factor $g_1^\pm$ in $M_1$

to $g_{1u}^\pm = \exp(i\mathbf{k} \cdot \mathbf{d}^{(1)} - i\pi\alpha) \pm \exp(i\mathbf{k} \cdot \mathbf{d}^{(1)'} + i\pi\alpha)$

and $g_{1d}^\pm = \exp(i\mathbf{k} \cdot \mathbf{d}^{(1)} + i\pi\alpha) \pm \exp(i\mathbf{k} \cdot \mathbf{d}^{(1)'} - i\pi\alpha)$, respectively.

The dimensionless magnetic field parameter is defined as $\alpha \equiv Bavb/(h/e)$. The in-plane field introduces the opposite phase shift in the upper (u) and lower (d) planes. So, the edge subband at the zigzag edge doubly splits in the horizontal direction as seen in Fig. 5(b). As seen in Fig. 5, the energy spectrum periodically changes with the period $\Delta\alpha = 1$ ($\Delta B = 1.17 \times 10^5$ T) in the ultra-high magnetic field region. It corresponds to the Aharonov-Bohm effect when the magnetic flux penetrates an atomic loop with the area of $(a/2) \times (2vb)$. Note that we did not take into account the Zeeman spin splitting.

As mentioned above, the qualitative features of phosphorene systems could be understood even by the simple LCAO model. In below, we review the one-year progress of phosphorene research mainly in physical properties.

### 3. Crystal Structure

Phosphorus exists as several allotrope forms.[13, 14] Among them, it is known that black phosphorous (BP) is most stable at room temperature. The crystal structure of bulk BP is displayed in Fig. 1(c). As confirmed in this figure, BP forms layered structure of two dimensional BP sheets, stacking parallel to crystallographic $z$-direction. The lattice constants of bulk BP are known as $a = 3.3136$ Å ($x$ direction), $b = 10.478$ Å ($z$ direction), and c = 4.3763 Å ($y$ direction).[15] Few-layer BP is usually obtained by mechanical exfoliation method from bulk BP, as in the case of graphene and other atomic layer materials. In addition, other exfoliation methods such as liquid exfoliation are also performed.[16, 17]

The crystal structure of monolayer phosphorene is shown in Fig. 1(a) and (b). Each phosphorus atom is covalently connected to three adjacent phosphorus atoms. As indicated in the figure, BP layers have armchair and zigzag directions, which correspond to crystallographic $x$- and $y$-direction, respectively. Unlike planar or buckled honeycomb structures in graphene and silicene, respectively, BP forms puckered honeycomb structure. The lattice constants of 1-5 layer phosphorene are estimated using density-functional theory (DFT) calculation. According to this calculation, it is suggested that with decreasing the layer numbers, the lattice constant parallel to armchair direction significantly increases, whereas that parallel to zigzag direction decreases.[18] Such extension of the lattice constant parallel to the armchair direction is mainly derived from an increase of the bond angle.[18]

Due to the unique puckered honeycomb structure, phosphorene shows anisotropic natures of electronic and phonon dispersions, which results in anisotropic electric-transport, thermal transport and optical properties as introduced in section 4, 8, and 6, respectively. Moreover, the puckered honeycomb structure leads to characteristic mechanical properties, especially significant flexibility against tensile and compressive strains.[19] In general, two-dimensional materials such as graphene and transition-metal dichalcogenides (TMDs) show superior mechanical flexibility. Compared with them, phosphorene is especially flexible material. The mechanical flexibility of phosphorene is theoretically examined in Ref. 20, and demonstrated that phosphorene can withstand stress up to 30%. Such elastic nature indicates the potential of phosphorene as flexible devices. Moreover, the high flexibility of phosphorene leads to the possibility of strain engineering. As introduced in Section 4, it is expected that the band structure of phosphorene can be tuned by applying moderate tensile or compressive strains and shows band convergence at the critical strain, which leads to drastic change of the anisotropy of the transport properties and an anomalous enhancement of the Seebeck coefficient.

Since phosphorus atoms in BP have free lone pairs and valence bond angle of 102°, BP shows lower chemical stability.[21] For example, bare BP reacts with $H_2O$ irreversibly and is easily degraded in ambient condition.[22] Therefore, it is essential to protect BP surface from atmosphere. In order to prevent from the degradation, the encapsulation by atomic layer deposited $AlO_x$ overlayer is suggested to be effective. Actually, it is confirmed that the encapsulated BP FET devices maintain their performance for more than a few months in ambient condition.[22, 23] The detail of the encapsulated FETs performance is described in Section 7.

## 4. Electronic structures

In Section 2, the electronic structure of phosphorene systems has been briefly considered using a simple LCAO model. Here, we review its more accurate studies. The band structure of monolayer phosphorene was first considered in 1981 using the tight-binding calculation[4]. The tight binding (or LCAO) model cannot reproduce the accurate band dispersion, since it takes into account only finite number of the Fourier components. Nevertheless, it has still an availability easy to interpret the results

physically.[24]

Most studies on electronic structure of phosphorene or thin BP are based on the first principle calculations.[1, 18, 19, 25–37] Their electronic properties are determined mainly by the band dispersions in the vicinity of Γ point. For example, anisotropic properties arises from the difference between Γ-X and Γ-Y dispersions. Here, Γ-X and Γ-Y directions correspond to armchair and zigzag directions, respectively. According to the first principle calculations, a direct band gap locates at Γ point for monolayer phosphorene, and it reduces with increasing the number of layers, reaching to 0.3 eV at Z point for bulk BP.[18, 24, 26] In this section, we describe the anisotropic band dispersions and the band gap in pristine and adatom-adsorbed samples. Subsequently, the electronic structure in nanoribbons is described.

Reflecting the anisotropic crystal structure, the band structure is anisotropic between Γ-X and Γ-Y directions. The effective mass along Γ-X direction (armchair direction) is lighter than that along Γ-Y direction (zigzag direction).[18, 19, 25] In the case of the monolayer phosphorene, the ratio of the mass between along armchair ($m_x$) and along zigzag ($m_y$) is calculated as $m_y/m_x \approx 6.6$ for electron and 42 for hole.[18] With increasing the number of layers, these values become moderate.[18] The details of calculated effective masses for electron and hole are listed in Table 2 in Ref. 18. The spatial anisotropy of the electron effective mass depicts an approximate "8" shape, as can be seen in Fig. 3 in Ref. 25. It is noteworthy that the mass anisotropy can be reversed with an application of ∼ 5% tensile strain or more.[25] The reversal mass anisotropy is derived from the band convergence.[19, 28, 29] Since the dominant orbitals and bond status (bonding, nonbonding, and antibonding) of each band are different, the strain dependence of each band is different.[19] For example, the energy level of the band with bonding nature descends with compression, while that with antibonding increases. As a result, a lowest-energy conduction band or a highest-energy valence band is altered to another band under a few percent of strain, namely, the band convergence occurs.

The values of the band gap $E_g$ are calculated by many groups,[1, 18, 19, 26, 27] while they depend on the employed calculation methods. The density functional theory (DFT) method gives the value of around 1 eV,[1, 19] but this method tends to underestimate $E_g$. It is widely accepted that the reliable $E_g$ is provided by the GW approach (GWA), which results in $E_g \approx 2$ eV.[26] The scanning tunneling microscopy/spectroscopy (STM/STS) supports this prediction of $E_g \simeq 2$ eV.[30] As discussed in Ref. 26, a strong band anisotropy, i.e., quasi-one-dimensional band dispersions cause stronger many-electron effect, leading to an enhanced $E_g$ from 0.8 eV to 2 eV. The difference between DFT and GW calculations is brought by this many-body effect.[26] In Ref. 26, the authors also discuss the exciton binding energy (around 800 meV), which reduces $E_g$. Although the absolute value of the gap is underestimated in DFT calculations, this method is useful enough to estimate the trend of the strain effects and the layer-number dependence of the electronic properties.[18, 19] When considering layer-number dependence in phosphorene, the interlayer interaction (0.46 eV) plays an important role.[18] A bonding-like feature of the valence band is found in interlayer region, which means the wave function overlap is the origin of the interlayer interaction, and introduces the band dispersion along z-direction. As a consequence, an abrupt change is visible between monolayer and bilayer, and the stronger interlayer interaction at thicker system induces the reduced $E_g$.[18] This prediction is experimentally confirmed. In fact, the peak of photoluminescence (PL) signal centered at 1.45 eV (in the visible wavelength) with a width of ∼ 100 meV can be detected in monolayer phosphorene.[1] This is in contrast to that in bulk BP, which is observed in the infrared wave region. The band gap is predicted to be controlled also by strain. As mentioned above, the strain causes the drastic change of the band structure through the band convergence, which yields direct-indirect band-gap transition, as well as changing the size of the band gap.[19, 29, 31] For example, the tensile strain applied to the armchair direction first enhances $E_g$, but at a critical strain at 5-8%, the band structure shows transition from direct to indirect band gap and $E_g$ starts to decrease.

Adsorption of adatoms is also promising way to produce a variety of structural or electronic states. Although it is difficult to prevent BP from ambient degradation due to high adsorption energy (more than twice as high as that in graphene),[32] once we make adsorption in a controllable way, this material is useful for energy storage, catalysis, gas sensing, nanoelectronics, and so on. The computational study of adatoms on monolayer phosphorene is systematically performed, and demonstrated in Refs. 32–36. In Ref. 32, the adsorption effect of 20 kinds of adatoms on the electronic density of states (DOS) is studied. Phosphorene has several adsorption sites (labeled H, B, B1, T, *etc.*), and each adatom has a preferred site. Most of metal adatoms sit on the H-site, while Si, Ge, and P adatoms on the B-sites. The effect on DOS is also different depending on the kinds of adatoms, for example, alkaline-metal only shifts Fermi level in a rigid-band-like manner, and transition-metal drastically changes DOS in the whole energy range, accompanying magnetic moment. They pointed out that oxygen atoms strongly bind with phosphorous atoms at the T-site, forming P=O double bond and polarized state. Defects or substitution effects, similar to the adsorption effects, are also studied in Refs. 33–42.

Inspired by the graphene nanoribbons, the electronic structure of phosphorene nanoribbons is intensively studied through simulations. In order to clarify the size effect or the edge character, the scaling laws, namely the energy gap as a function of the ribbon width are examined.[43–54] The effects of passivation to the edges are also studied. The band gap becomes larger with reducing the width of nanoribbons, due to quantum confinement effect.[44, 49] We should note that the direction of cut of nanoribbons affects the electronic properties. According to Refs. 46, 48, armchair nanoribbons show semiconducting behavior with indirect band gap, while zigzag nanoribbons exhibit metallic behavior due to the emergence of the edge state. With a treatment of hydrogenation of the edge, both types of nanoribbons exhibit semiconducting behavior with a direct band gap. The deformation energy, effective mass, and mobility in several kinds of nanoribbons are summarized in Table 1 of Ref. 48. Besides, the novel magnetic state is proposed in zigzag phosphorene.[50, 51] Namely, the ground state of it shows ferromagnetic in a given edge, but antiferromagnetic order between two opposite edges. The origin of the magnetism is attributed to the dangling bonds and the localized edge $\pi$-orbital states.

As mentioned above, the edge state is significantly important to consider the transport and magnetic properties in phosphorene. The edge state spectrum can be obtained by calculating the energy levels in nanoribbons. It has been performed by using the first principle[54, 55] and tight binding[10, 56] calculations. As already mentioned in Section 2, there exists an isolated half-filled edge subband in the middle of main gap at the zigzag edge in monolayer, whereas there are fully occupied and empty edge subbands at the armchair edge. In graphene with a honeycomb lattice and two Dirac cones, it is well known that there exists a zero-mode edge state (Fujita state) connecting two Dirac cones at the zigzag edge.[57] The topological origin of the isolated edge state in phosphorene is discussed in relation to the edge state in graphene.[56] In phosphorene with a puckered honeycomb lattice, the edge state is left in the gap resulting from the merging of two Dirac cones.[56] Since the edge subband at the zigzag edge is half-filled, it could cause the metallic edge transport in monolayer. In bilayer, the edge states are overlapped with the valence band, so that the charge transfer occurs between the edge state and the valence band. This causes the local hole doping around the zigzag edge.[10]

## 5. Phonon structures

As in the case of the electronic structure, phonon structure of phosphorene also shows significant anisotropic nature. Since the unit cell of monolayer phosphorene contains four phosphorus atoms, there are 12 phonon modes in total, namely three acoustic phonon modes and nine optical phonon modes. The optical mode at the Γ point can be classified into $B_{1u}$, $B_{2u}$, $B_{2g}$, $2B_{3g}$, $2A_g$, $A_u$, and $B_{1g}$ due to its orthorhombic symmetry. All the even parity modes are Raman active, whereas the odd $B_{1u}$ and $B_{2u}$ modes are infrared (IR) active. As introduced in Section 6 in detail, the Raman spectra can be a powerful tool for sample characterization such as sample orientation and the strength of strain applied for phosphorene.[58] On the other hand, three acoustic phonon mode can be classified into two in-plane acoustic modes (the transverse acoustic mode (TA) and the longitudinal acoustic mode (LA)) and the off-plane acoustic mode (ZA). The two in-plane acoustic modes exhibit linear dispersions, while the off-plane acoustic mode (ZA) exhibits a parabolic dispersion.[59] Theoretical calculations indicate that acoustic phonon modes exhibit anisotropic nature due to its puckered honeycomb structure, which results in the anisotropic sound velocity of LA mode parallel to armchair ($\Gamma$-X) direction and that parallel to zigzag ($\Gamma$-Y) direction. Namely, the sound velocity of the LA branch parallel to zigzag direction is larger than that parallel to armchair direction. The difference results in the anisotropic nature of thermal conductivity, which is discussed in Section 8.

## 6. Optical properties

Reflecting the anisotropic band structures, optical absorption spectra show linear dichroism for incident light along the $z$ direction and linearly polarized in the $x$ (armchair) and $y$ (zigzag) directions.[18] According to Ref. 18, the strong layer number dependence of the first absorption peak is expected theoretically. For a dielectric polarization in the $x$ direction, the peak is found at $1.55$ eV in monolayer phosphorene, and with increasing layer number, it monotonically reduces, and reaches to $0.46$ eV in bulk BP. For $y$-polarized light, on the other hand, only a slight decrease is found from $3.14$ eV in monolayer to $2.76$ eV in bulk BP with increasing layer number. Another result is that, from the infrared (IR) to visible light range, BP strongly absorbs the light polarized along the $x$ direction, but it is transparent to the light polarized along the $y$ direction. This is because the dipole operator connects the valence band and conduction band states for the $x$-polarization, but symmetrically forbidden for the $y$-polarization. Such an anisotropic optical absorption[27] can be utilized to determine the crystal axis. The crystalline orientation determined by the polarization-dependence of IR coincides with the result of the angle-dependence of the DC conductance,[1] which is described in Section

7. Raman spectroscopy is also a powerful tool for the characterization of the structural properties. Although Ref. 1 reported the dependence of the positions of the Raman peak $A_g^1$ and $A_g^2$ on the layer number, experimental consensus has not been reached yet.[60, 61] The divergent results are possibly due to ambient degradation[21] or strong temperature dependence.[61, 62] By contrast, as pointed out in Refs. 60, 62, a relative intensity of $A_g^1$, $A_g^2$, and $B_{2g}$ modes strongly depend on the polarization angle of incident light. This fact shows that the polarization dependent Raman spectroscopy is also useful for determining the crystal orientation. It should be noted that the relatively large temperature dependencies of the Raman shift is observed in phosphorene,[61, 62] with the slope of $\sim -0.02 \text{ cm}^{-1}/°C$, compared with that in bilayer graphene ($-0.0154 \text{ cm}^{-1}/°C$ for G peak) and few-layer MoS$_2$ ($-0.0123$ and $-0.0132 \text{ cm}^{-1}/°C$ for $A_{1g}$ and $E_{2g}^1$ modes, respectively).[62] The larger temperature dependence reflects the mechanical flexibility of phosphorene. In addition, it is suggested that the Raman spectroscopy is useful to investigate the strain introduced in phosphorene.[58] The optical properties (IR and Raman) enable us to instantly determine the crystal orientation or layer number without a help of transmission electron microscope (TEM) or scanning tunneling microscope (STM).[62]

**7. Transport properties**

Transport properties in few-layer BP (multilayer phosphorene) have been investigated after the fabrication of the field-effect transistor (FET) devices. Detailed descriptions are found in Ref. 63 for experimental techniques to fabricate and characterize BP FETs. Few-layer BP FETs were firstly reported in Refs. 1 and 2, where BP was transferred onto a SiO$_2$-coated silicon wafer using a mechanical exfoliation method in the same way as graphene. Metal contacts such as Ti/Au were deposited after the standard electron-beam lithography process. A typical thickness of the BP in the devices was $5 \sim 10$ nm or more, which was determined by the atomic force microscope (AFM) measurements. Taking into account an adjacent layer spacing of $\sim 0.53$ nm for monolayer phosphorene,[60, 64] the layer number of the devices corresponds to the order of 10. Unfortunately, the thinner samples were rarely succeeded in measuring transport properties. Their ambient degradation nature may be one of the reasons to prevent monolayer phosphorene from the transport measurement. The exception is found in Ref. 65, where monolayer, bilayer, and trilayer phosphorenes are successfully measured with the structure sandwiching phosphorene with hexagonal boron nitride (h-BN). The crucial effect of h-BN sandwich is elaborated at the end of this section. Monolayer phosphorene FETs are studied only from theoretical approach, as found in Ref. 66. For the purpose to keep it away from such degradation, encapsulation with alumina (Al$_2$O$_3$) is proved to be effective.[21–23, 67–69] For example, in Ref. 21, an Al$_2$O$_3$ coating followed by a hydrophobic fluoropolymer film keeps the FETs performance for at least a few months. Annealing effect is also investigated combined with such encapsulation.[23, 70] Several kinds of materials for encapsulating overlayers are theoretically explored in terms of the charged impurity-limited carrier mobility,[68] resulting in an enhanced mobility by up to an order of magnitude with high-$\kappa$ overlayer, such as Al$_2$O$_3$ ($\kappa = 12.5$) and HfO$_2$ ($\kappa = 22$). This is similar to the case of TMDs. The origin of the enhancement is dielectric screening of the charged impurities.

Due to the anisotropic mass, the transport properties are also expected to be anisotropic, in contrast to other conventional semiconductors. Experimentally, the first attempt to estimate the anisotropy was performed by using an angle-resolved DC conductance measurement, where every 30° or 45°-spaced electrodes were attached onto a thin sample, and performed DC conductance measurements across each pair of diagonally positioned electrodes.[1, 60] Around 50% larger conductivity in the armchair ($x$) direction than that in the zigzag ($y$) direction is clearly observed, in spite of some tens of angular resolution. Similar but slightly larger anisotropy is found in Hall mobility measurements.[60] As pointed out in Ref. 1, the anisotropy of the mobility is possibly underestimated since the current spreading might partially average out the anisotropy. However, we should keep in mind that not only the mass anisotropy determine the anisotropy of the mobility.[18] The carrier mobility in a phonon-limited scattering model (intrinsic limit) is calculated with the expression

$$\mu_{2D} = \frac{e\hbar^3 C_{2D}}{k_B T m_e^* m_d (E_i^i)^2},$$

(6)

where $m_e^*$ is the effective mass in the transport direction and $m_d = \sqrt{m_x^* m_y^*}$ is the average effective mass, $E_i^i = \Delta V_i/(\Delta l/l_0)$ is the deformation potential constant of the $i$-th band along the transport direction, and $C_{2D}$ is the elastic modulus defined by $(E - E_0)/S_0 = C(\Delta l/l_0)^2/2$. $e$, $\hbar$, $k_B$, and $T$ are elementary charge, Planck constant divided by $2\pi$, Boltzmann constant, and temperature, respectively. According to this model, the mobility is predicted to be higher in the lighter-mass direction (armchair direction) in 2~5-layer phosphorene, in which the anisotropy of the mobility is $\mu_x/\mu_y \approx 1.5$~2. Thus, on-current along armchair direction is expected to be larger than that along zigzag direction.[71] Although the mobility in the realistic FETs are probably limited not by phonon scattering but by impurity scattering, the observed anisotropy of the mobility in few-layer BP FETs is in good agreement with the calculated value. The anisotropy assuming the charged impurity scattering is presented in Fig. 5 in Ref. 68. The experimental value of mobility is found, for example, in Ref. 60. At a constant hole doping concentration of $6.7 \times 10^{12} \text{cm}^{-2}$ by tuning $V_g$, the mobility at room temperature shows $400{\sim}600 \text{ cm}^2/\text{V} \cdot \text{s}$, and reaches the highest value of $1000 \text{ cm}^2/\text{V} \cdot \text{s}$ at $120 \text{ K}$ for the armchair ($x$) direction. An asymmetric mobility of hole and electron is observed in electric-double-layer transistors (EDLT),[72] which is consistent with theoretical prediction.[18] These values of mobility are still smaller than those in bulk BP, where the electron and hole mobility is $\sim 1000 \text{ cm}^2/\text{V} \cdot \text{s}$ at room temperature, and it could exceed $15000 \text{ cm}^2/\text{V} \cdot \text{s}$ for electrons and $50000 \text{ cm}^2/\text{V} \cdot \text{s}$ for holes at low temperatures.

In order to characterize the semiconducting properties in the FETs device, the switching property was examined by applying the gate voltage $V_g$ to the heavy-doped silicon gate under a fixed source-drain voltage. A measured source-drain current shows a sizable enhancement by a factor of $\sim 10^5$ when $V_g$ varies from zero to negative (typically $-30 \text{ V}$), i.e. *p*-type behavior.[2, 60, 73] Note that some devices with a higher mobility show ambipolar behavior, namely, a moderate enhancement of current can be seen also in $V_g > 0$. The choice of the metal also affects the ambipolar character, as shown in Refs. 74, 75. In Refs. 66, 74, 76, it is discussed that the ambipolar character can be controlled also by the channel length, as well as the work function of the metal. They fabricated FETs with various channel lengths, and discuss the effect of drain-induced barrier lowering (DIBL) on the appearance of the ambipolar character. A modulation of the drain current is four orders of magnitude larger than that in graphene due to the presence of a band gap, and comparable to MoS$_2$ devices. Besides, a *p*-type behavior indicates a possible counter material for TMDs, which shows either *n*-type or ambipolar as a consequence of the energy level of S vacancy and charge-neutral level near the conduction band edge in these materials. We should comment that the on-off current ratio strongly depends on the thickness of BP. For a sample with ∼2 nm in thickness, the on-off ratio at room temperature is as high as $5 \times 10^5$, while that in a sample with thickness > 15 nm is reduced to less than 10.[60] A thinner BP device is desirable for future applications. We also note that transport-direction dependent current characteristics suggests possible ballistic device performance in monolayer phosphorene.[71] Namely, a higher on-state current along the armchair direction would be realized in monolayer. The field-effect mobility $\mu_{\text{FE}}$ is also estimated in these devices from the expression

$$\mu_{\text{FE}} = \frac{G_m}{C_{\text{OX}}(W/L)V_{\text{ds}}},\tag{7}$$

where $C_{\text{OX}}$ is the capacitance of the gate oxide, $W$ and $L$ are the channel width and length, $V_{\text{ds}}$ is the drain bias, and $G_m = dI_{\text{ds}}/dV_g$ is the trans-conductance ($I_{\text{ds}}$ is the drain current). The obtained values at room temperature is $200 \sim 300 \text{ cm}^2/\text{V} \cdot \text{s}$, even in intentionally aligned samples. Although a clear difference is not recognized between intentionally aligned devices and other samples for the case of few-layer BP devices so far, such an alignment would give a possible chance to improve the device performance when the monolayer phosphorene FETs are successfully fabricated.[71]

In fabricating the FETs, the metal contacts significantly affect to their behavior.

Depending on the difference of work function between BP (semiconductor) and contact metal (Ti, for example), the junction behaves as Ohmic or Schottky barrier.[77] The relative position between the bands of BP and the Fermi level of the metals is illustrated in Fig. 5 in Ref. 78. According to the first-principles study, several metals are found to have minimal lattice mismatch with monolayer phosphorene, and Cu(111) is the best candidate to form Ohmic contact.[79] Experimentally, Ti and/or Au contacts are often employed, resulting in the Schottky contacts.[60, 70, 75] The Schottky barrier height, as large as 0.21 eV, is determined by the three terminal method in Ref. 1. It can be tuned by the gate voltage $V_g$.[67]

Since the band gap can be controlled by the thickness or strain, this material can be used as a tunable photodetector. In fact, a photocurrent was experimentally confirmed with short response time (typically 1 ms for a rise time).[80] The superiorities of BP photodetectors are sizable responsivity (4.8 mA/W), as well as broad spectral range from visible to near infrared wavelength. Here responsivity is defined as $I_{ph}/P_{laser}$, where $P_{laser}$ is the incident power of the laser, and $I_{ph}$ is the increase of the drain current by laser illumination. The responsivity can be enhanced in device structures with *pn* junctions. The vertical *pn* junction was realized in a heterojunction of few-layer BP and monolayer MoS$_2$, in which a maximum value of 3.54 A/W is confirmed.[81, 82] The horizontal *pn* junction was formed by using a split gate, which works on a local area of few-layer BP, without a help of TMDs.[83] Similar effect is also expected in the Schottky junction. Photocurrent signal is detected only around the metal contact of the FETs, exhibiting the function of the photodetector resides in the Schottky junction.[73] In other devices, different origins of the photocurrent are discussed, for example, photovoltaic, photothermoelectric, and photobolometric effects.[84] Some other prototype FETs are already fabricated, such as radio-frequency FETs,[85] flexible amplitude-modulated (AM) demodulator.[69]

Before the end of this section, we stress the advantage of BP/h-BN heterostructures. Recently, the structure is recognized as a powerful way to realize high-quality device.[65, 86, 87] Ref. 87 discusses the effect of the underlying substrate, which affects the transport properties in the similar way to graphene and TMDs. In fact, phosphorene on h-BN drastically improves the ambipolar property,[65, 72, 86–89] whereas the encapsulation effect is thought to be also important. In the high-quality samples on h-BN, the quantum oscillation is observed.[86, 88, 89] In addition, it is noteworthy that the integer quantum Hall effect is successfully observed.[88]

## 8. Thermal properties

It is known that graphene has a high thermal conductivity, which is up to 5000 W/K·m.[90] Such high thermal conductivity of graphene is originated from the acoustic ZA mode, which corresponds to the out-of-plane vibration mode. The contribution of the ZA mode to the thermal conductivity of graphene reaches up to 75 %.[91] On the other hand, the thermal conductivity of phosphorene is theoretically estimated to be much smaller than that of graphene and TMDs. Such low thermal conductivity is derived from a large scattering rate of the ZA mode due to the puckered structure.[92, 93]

In addition to the lower thermal conductivity of phosphorene, it is also indicated that the thermal conductivity of phosphorene is highly anisotropic, in contrast to isotropic one in graphene and TMDs.[59, 94, 95] It is expected that the thermal conductivity parallel to zigzag direction is much larger than that parallel to armchair direction. Such anisotropic nature of the thermal conductivity is originated from the anisotropic phonon dispersion due to its puckered structure. From theoretical analyses, it is indicated that the sound velocity along zigzag direction is about twice larger than that along armchair direction, which leads to the anisotropic thermal conductivity.[59, 94, 95] On the other hand, it is also suggested that the anisotropy becomes considerably small in the ballistic regime in Ref. 96.

The strain effect against the thermal conductivity is investigated in Ref. 94. It is indicated that in the case of applying biaxial strain and uniaxial strain parallel to armchair direction, the thermal conductivity of both zigzag and armchair direction is reduced. However, in the case of applying uniaxial strain parallel to zigzag direction, the thermal conductivity parallel to ziagzag direction is enhanced while armchair direction is decreased. Such anisotropic strain dependence of the thermal conductivity is due to the enhancement of the contribution of the low frequency phonon modes. These theoretical results indicate the possibility to modulate the thermal conductivity of phosphorene by strain.

Hereafter, the details of the thermoelectric properties of phosphorene is introduced. In the theoretical study, it is suggested that phosphorene is a promising thermoelectric material. The electrical conductivity and thermal conductivity show in-plane anisotropy and the preferred direction of electrical conductivity and thermal conductivity is orthogonal to each other. The preferred direction of electrical conductivity is parallel to armchair direction, although that of thermal conductivity is parallel to zigzag direction. Such orthogonal relation can lead to the enhancement of the efficiency of the thermoelectric conversion. In general, the efficiency of thermoelectric materials are estimated by their figure of merit

$$ZT = \frac{S^2 \sigma T}{\kappa}$$

, (8)

where $S$ is Seebeck coefficient, $\sigma$ is electrical conductivity, $\kappa$ is total thermal conductivity of electron and phonon, and $T$ is temperature. For realistic application, high $ZT$ value, at least $ZT > 1$, is desirable. As confirmed in the equation, the enhancement of $\sigma/\kappa$ value leads to the improvement of the thermoelectric conversion efficiently. Therefore, large $\sigma$ and small $\kappa$ materials are desirable for realizing large $ZT$ value. In the case of phosphorene, the orthogonal relation of the preferred electrical and thermal conductivity can lead to the great enhancement of the thermoelectric performance. Actually, the in-plane direction dependence of the $ZT$ value of phosphorene is theoretically examined in detail and it is indicated that by optimizing the sample alignment and carrier concentration by tuning gate voltage, phosphorene can satisfy the criterion $ZT > 1$ even at room temperature.[59]

We introduce the possible enhancement of Seebeck coefficient by strain. As introduced in Section 4, phosphorene exhibits novel band convergence by applying strain. The band convergence leads to significant enhancement of Seebeck coefficient and power potential expressed as $P = S^2\sigma$. In conventional semiconductors, Seebeck coefficient depends on the density-of-state effective mass. Therefore, large number of degenerated valleys in the band structure can lead to the large Seebeck coefficient, resulting in the high thermoelectric performance. Actually, some representative thermoelectric materials such as Bi$_2$Se$_3$ are multivalley systems. In the case of phosphorene without applying strain, the band structure has a simple direct gap and there are no degenerated conduction and valence bands. However, by applying strain, the energy levels of several bands are gradually changed. As a result, the energy levels of conduction or valence band edges are converged at critical strain as indicated in Refs. 28, 97. Such band converged situation leads to the drastic enhancement of effective mass, resulting in the enhancement of Seebeck coefficient. As a result, significant enhancement of $ZT$ value is expected.

## 9. Other topics

The possibility of superconductivity in doped phosphorene was theoretically considered with Bardeen-Cooper-Schrieffer (BCS) mechanism.[98, 99] In Ref. 98, it is predicted that the superconductivity appears in doped monolayer phosphorene at electron density above $1.3 \times 10^{14}$ cm$^{-2}$, reflecting the dependence of the coupling constant on doping concentration. This superconductivity is also affected by the strain. The dominant electron-phonon coupling originates from the softening of the out-plane stretching mode ($A_g^1$) and the in-plane stretching mode ($B_{3g}^1$). The maximum critical temperature is expected to be higher than 10 K. Besides, in Ref. 99, BCS superconductivity is predicted in Li-intercalated bilayer phosphorene with the maximum critical temperature of 16.5 K. So far, the superconductivity is not experimentally confirmed yet even in EDLT devices.[72] In addition to superconductivity, it was theoretically proposed that the topological insulator phase could be induced under the electric field.[100] On the other hand, the possible nano-phosphorus structures, such as nanotube and fullerene, were also discussed.[43, 101]

Various applications of phosphorene systems have been proposed based on theirs advantageous properties. Using their high adsorption energy, the applications to a gas sensor,[102] a Li-ion battery anode,[103–105] and a hydrogen storage[106] are suggested. As nanoscale devices, a photodetector,[107] a radio-frequency transistor,[85] a nano-electromechanical resonator,[108] a flexible rewritable memory,[109] a tunnel FETs,[110] and a photocatalyst[111] are intensively studied.

## 10. Summary

In the first part, using a simple LCAO model, we have qualitatively demonstrated the characteristic features of electronic structures of phosphorene systems; the reduction of band gap by the layer number and the vertical electric field, the anisotropy of valence band dispersion with lighter hole mass along the armchair direction, the appearance of the midgap edge states at the zigzag edge, and their dependence on the vertical electric and in-plane magnetic fields. In the second part, we have reviewed the recent progress of phosphorene researches. In spite of their ambient degradation, phosphorene systems are semiconductors with various desirable properties; high mechanical flexibility, high adsorptive activity, high electron mobility, high on/off ratio in FETs, anisotropic transport, large figure of merit for thermoelectric conversion, etc. Using these properties, various applications have been considered.

**Acknowledgements**

The authors thank to Prof. Y. Akahama and Prof. H. Tajima for arousing our interest in phosphorene, and Dr. M. Ezawa, Prof. Y. Fuseya, and Prof. M. Tokunaga for valuable discussions. This work is supported by a Grant-in-Aid for Scientific Research on Innovative Areas "Science of Atomic Layers" (No. 25107003) from the Ministry of Education, Culture, Sports, Science, and Technology, Japan.

**Figure 1**

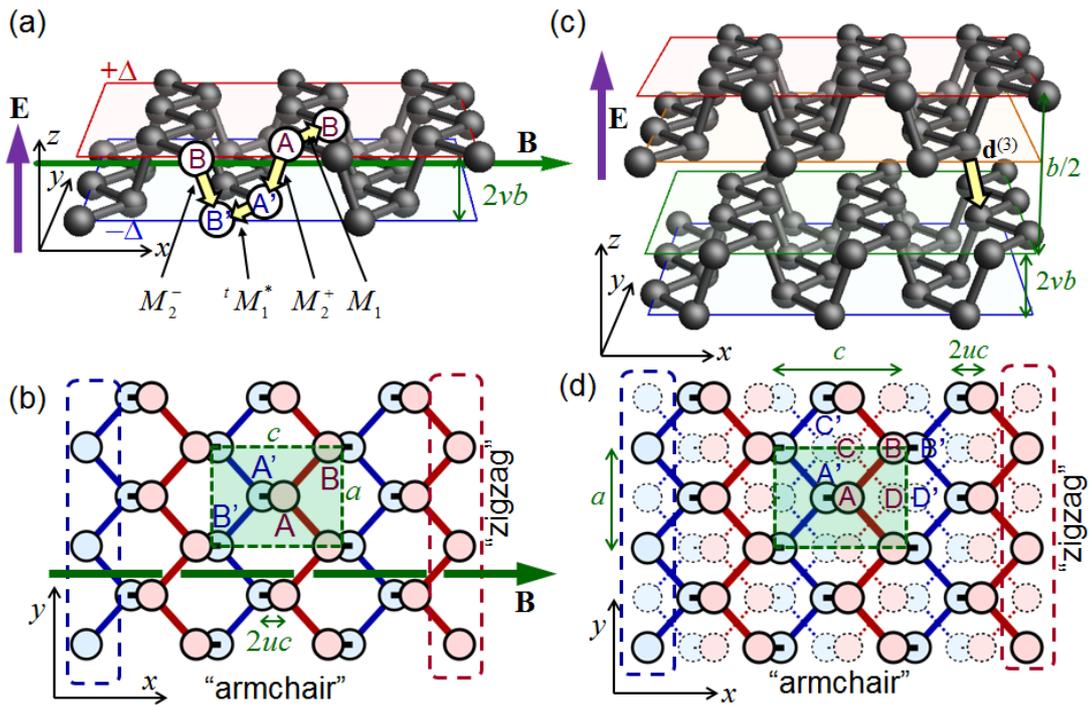

**Fig. 1.** (color online)

(a) Crystal structure of monolayer phosphorene. Inter-atomic coupling matrices are indicated. (b) 2D lattice of monolayer projected onto the *xy*-plane. Dashed rectangle indicates the unit cell. The "zigzag" and "armchair" edges are also indicated. (c) Crystal structure of bilayer phosphorene. (d) 2D lattice of bilayer projected onto the *xy*-plane.

**Figure 2**

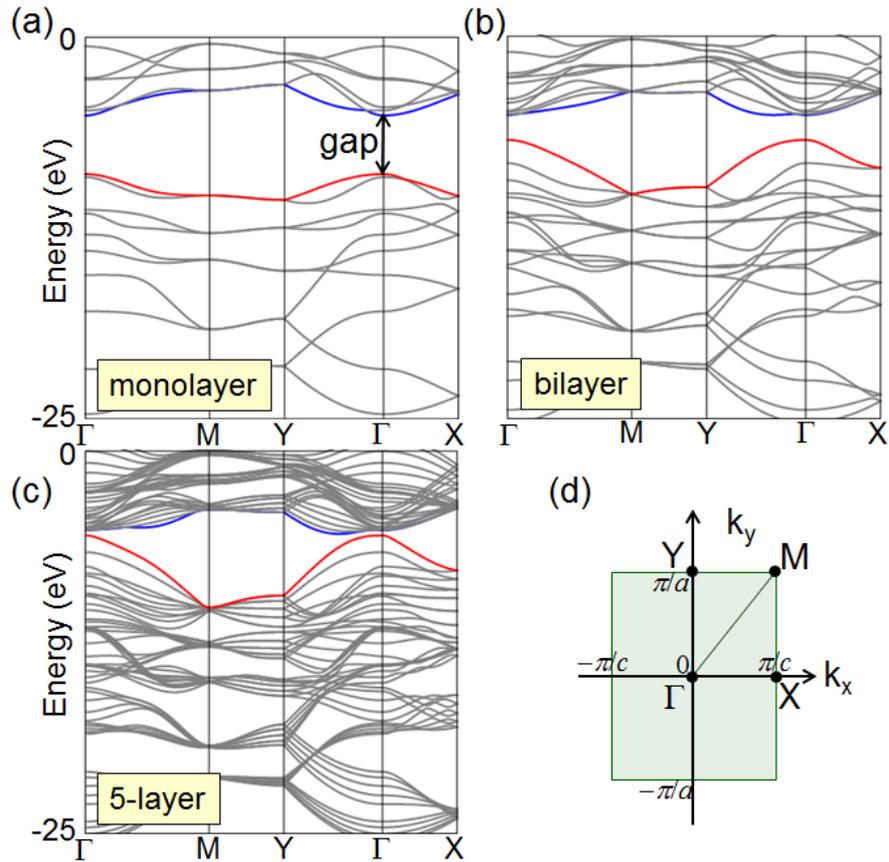

**Fig. 2.** (color online)

LCAO band dispersion of (a) monolayer, (b) bilayer, and (c) five-layer phosphorene. Only the nearest neighbor and next nearest neighbor in-plane couplings and the nearest neighbor interlayer coupling are taken into account. (d) The symmetric points in the 2D Brillouin zone.

**Figure 3**

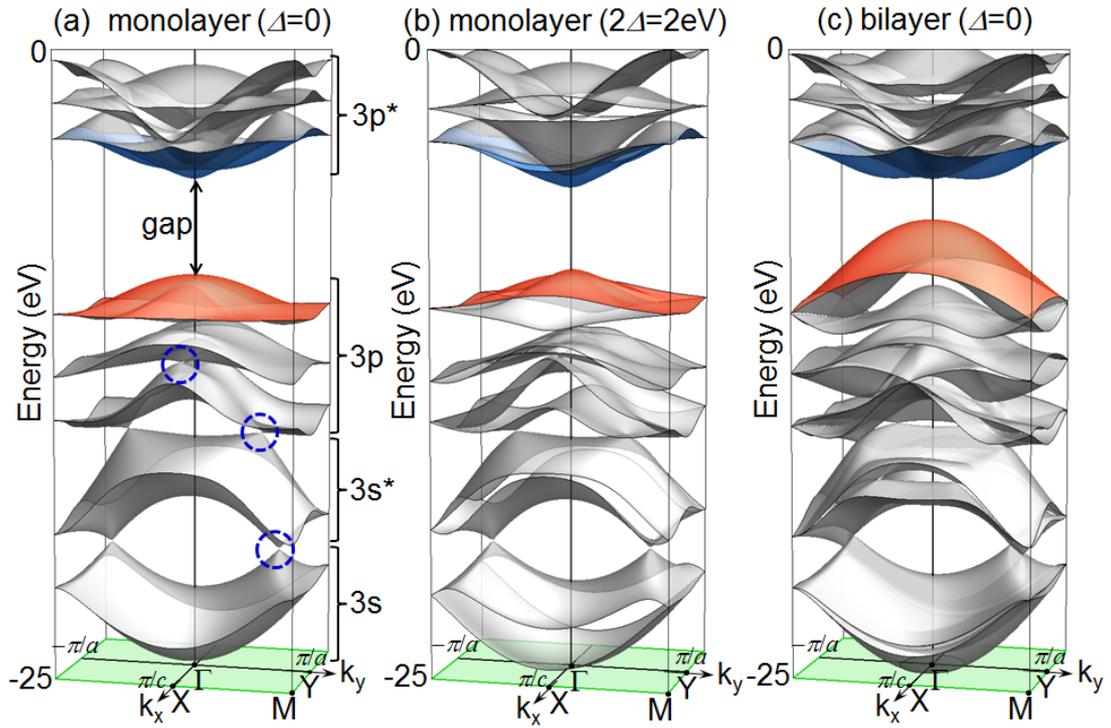

**Fig. 3.** (color online)

Global band structure of (a) monolayer phosphorene, (b) monolayer under the vertical electric field, and (c) bilayer phosphorene. Several Dirac points are indicated by dashed circles.

**Figure 4**

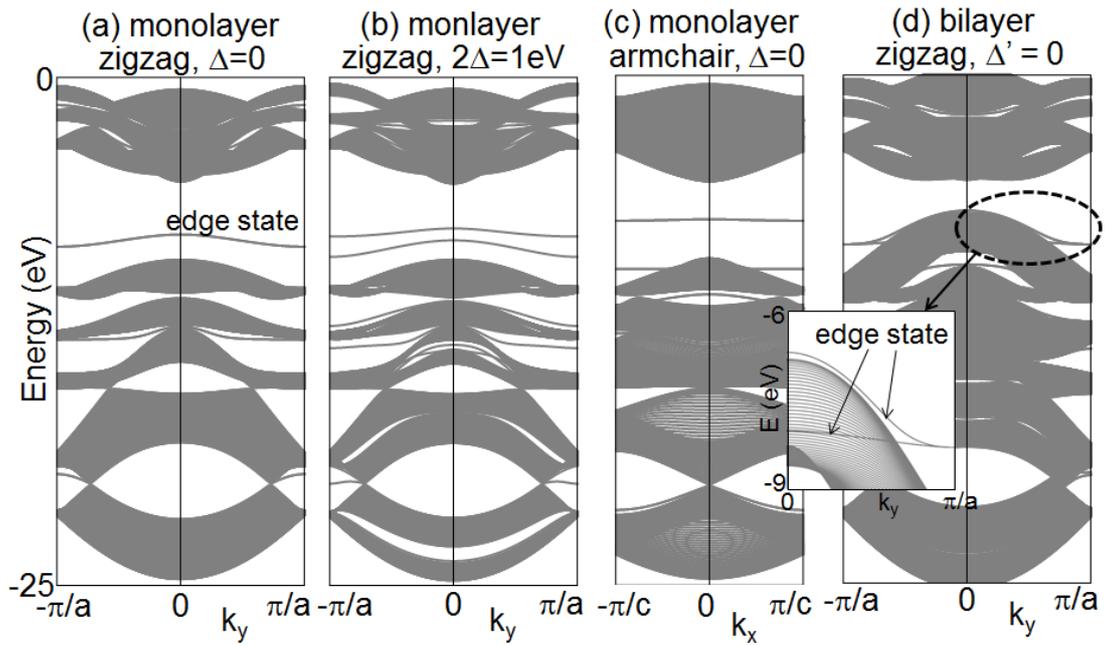

**Fig. 4.**

Energy spectra of phosphorene nanoribbons. (a) Monolayer nanoribbon with the zigzag edge. (b) Monolayer nanoribbon under the electric field. (c) Monolayer nanoribbon with the armchair edge. (d) Bilayer nanoribbon with the zigzag edge. The inset shows the details of the edge states in the bilayer nanoribbon.

**Figure 5**

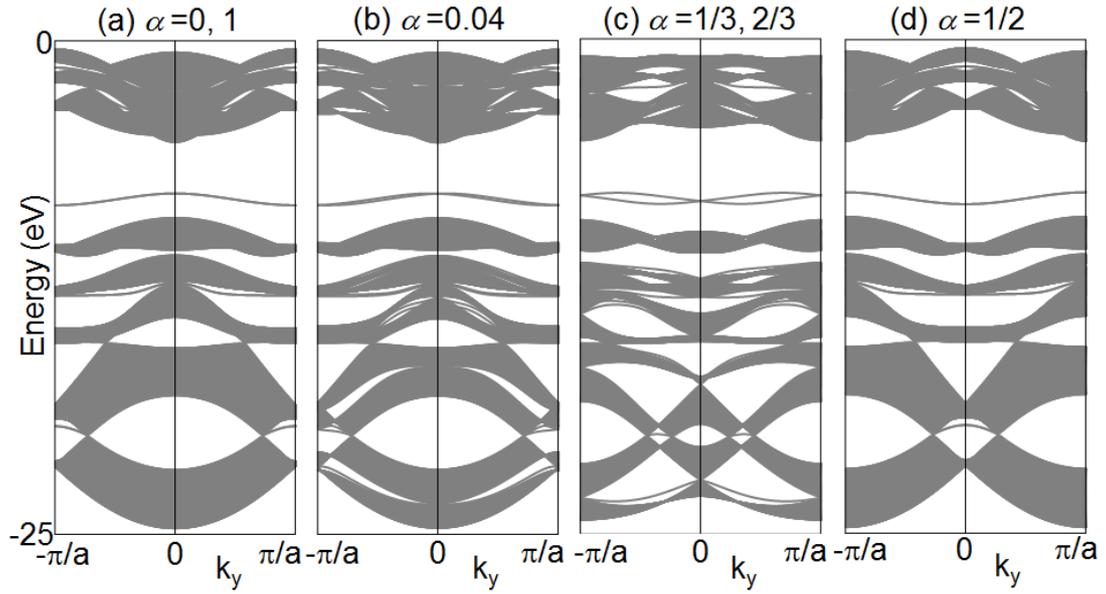

**Fig. 5.**

Energy spectra of monolayer phosphorene nanoribbon with the zigzag edge under the in-plane magnetic fields (a) $\alpha = 0$, $\alpha = 1$, (b) $\alpha = 0.04$, (c) $\alpha = 1/3$, $\alpha = 2/3$, and (d) $\alpha = 1/2$. The magnetic field is applied normal to the zigzag edge.